\documentclass[12pt]{article}
\usepackage{epsfig}

\begin{document}
\def\br{\begin{eqnarray}}
\def\er{\end{eqnarray}}
\def\be{\begin{equation}}
\def\ee{\end{equation}}
\def\({\left(}
\def\){\right)}
\def\a{\alpha}
\def\b{\beta}
\def\d{\delta}
\def\D{\Delta}
\def\g{\gamma}
\def\G{\Gamma}
\def\h{ {1\over 2}  }
\def\hp{ {+{1\over 2}}  }
\def\hm{ {-{1\over 2}}  }
\def\k{\kappa}
\def\l{\lambda}
\def\L{\Lambda}
\def\m{\mu}
\def\n{\nu}
\def\o{\over}
\def\O{\Omega}
\def\p{\phi}
\def\rh{\rho}
\def\s{\sigma}
\def\t{\tau}
\def\th{\theta}
\def\ii {\'\i  }

\title{{\small {\bf DEFORMED WOODS$-$SAXON
POTENTIAL IN THE FRAME OF SUPERSYMMETRIC QUANTUM MECHANICS FOR ANY
$l$-STATE }}}

\author{C\"uneyt Berkdemir $\thanks{{\small {\sl Permanent
Address: Department of Physics, Faculty of Science and Literature,
Erciyes University,}} {\small {\sl 38039, Kayseri, Turkey}}}$\\
%EndAName
{\small {\sl The Abdus Salam International Center for Theoretical
Physics,
Trieste, Italy}}\\\\
Ay\c se Berkdemir \\
%EndAName
{\small {\sl Department of Physics, Faculty of Science and
Literature, Erciyes University, 38039, Kayseri, Turkey}}\\\\
and \\\\
Ramazan Sever $\thanks{{\small \sl Corresponding author:
sever@metu.edu.tr}}$ \\
%EndAName
{\small {\sl Department of Physics, Middle East Technical
University,}} {\small {\sl 06531, Ankara, Turkey}}}
\date{}

\maketitle

\begin{abstract}
A novel analytically solvable deformed Woods$-$Saxon potential is
investigated by means of the Supersymmetric Quantum Mechanics.
Hamiltonian hierarchy method and the shape invariance property are
used in the calculations. The energy levels are obtained for any
$l$-state.
The interrelations for some nuclear scattering processes are also discussed.\\
\end{abstract}

\baselineskip=22pt plus 1pt minus 1pt
%%%%%%%%%%%%%%%%%%%%%%%%%%%%%%%%%%%%%%%%%%%%%%%%%%%%%%%%%%

\vspace{-0.5cm}



\vspace{0.6cm} \baselineskip=18pt
%\vfill
\begin{center}
MIRAMARE -- TRIESTE\\
\smallskip

February 2005\\
\end{center}
%\vfill
\vspace{1.0cm}

\newpage

\section{Introduction}
Supersymmetric quantum mechanics (SQM for short) is a framework
used to determine energy eigenvalues and eigenfunctions of quantum
mechanical problems. SQM appeared 25 years ago and has so far been
considered as a new field of research, providing not only a
supersymmetric interpretation of the Schr\" {o}dinger equation,
but important results to a variety of non-relativistic quantum
mechanical problems \cite{Cooper1}. One of the most important
works is that Gendenshtein introduced a discrete parametrization
invariance labelled "shape invariance" \cite{Gendenshtein}. For
the shape invariance potentials, their exact energy levels can be
found analytically by making use of the supersymmetry shape
invariance approach \cite {bar, balan, Levai}.

The association of the factorization and the hierarchy of
hamiltonian method with SQM formalism has been introduced to
obtain the approximate energy spectra of non-exactly solvable
potentials, \cite{Gozzi, Drigo4} as well as the partially
solvable, \cite{Drigo1, Drigo3}, the isospectral, \cite{Dunne},
the periodic potentials, \cite{Sukhatme} and the exponential-type
potentials \cite{bulent, sheng, metin, fakhri}. Using the physical
arguments, it is possible to make an {\it ansatz} for the
superpotential which satisfies the Riccati equation by an
effective potential. Thus, one can obtain a solution of the
Riccati differential equation for the superpotential function.
This new scheme can be successfully applied to obtain the spectra
of the nuclear systems well fit by the deformed Woods-Saxon
potentials.

In this work, a study of such new exactly solvable effective
potential through SQM is presented by the hierarchy of hamiltonian
method for $l=0$ and $l\neq 0$ cases. We first introduce SQM
briefly, and then show the deformed Woods-Saxon potential
hierarchy and its shape invariance property for the $l=0$ states.
The deformed Woods-Saxon potential hierarchy is recovered with
certain potential constants. The results obtained are improved by
means of the shape invariance for the $l\neq 0$ states. In this
case, the effective potential behaves in the same way  as
centrifugal part of the Woods-Saxon potential under the condition
of the first approximation. Finally, this
approximation is applied to the nuclear scattering process. \\

\section{Supersymmetric Quantum Mechanics}

The simplest way of generating a new exactly solvable Hamiltonian
from a known one is just to consider an invertible bounded
operator. In this case, a given Hamiltonian and its supersymmetric
partner have identical spectra except the zero energy of ground
state. In Supersymmetric Quantum Mechanics (SQM) for $N=2$ we have
two nilpotent operators, $Q$ and $Q^+$, satisfying the algebra \be
\{ Q, Q^+\} = H_{S} \;,\;\;\; \{ Q, Q\}= \{ Q^+, Q^+\} = 0, \ee
where $H_{S}$ is the supersymmetric Hamiltonian. The fact that the
supercharges $Q$ and $Q^+$ commute with $H_S$ is responsible for
the dejeneracy. This algebra can be realized as \be Q = \left(
\begin{array}{cc} 0  & 0  \\ A^-  & 0
\end{array} \right ) \;,\;\;\;
Q^+ = \left( \begin{array}{cc} 0  & A^+  \\ 0 & 0
\end{array} \right )
\ee where $A^{\pm}$ are bosonic operators.  With this realization
the supersymmetric Hamiltonian $H_{S}$ is given by \be H_{S} =
\left( \begin{array}{cc} A^+A^-  &  0 \\ 0 & A^-A^+
\end{array} \right ) = \left( \begin{array}{cc} H^-  &  0 \\ 0 & H^+
\end{array} \right ).
\ee $H^{\pm}$ are called supersymmetric partner Hamiltonians and
share the same spectra, apart from the nondegenerate ground state,
(see \cite{Cooper1} for a review), \be E_n^{(+)} = E_{n+1}^{(-)}.
\ee For the non-spontaneously broken supersymmetry this lowest
level is of zero energy, $E_1^{(-)} = 0$. We have \be
H^{\pm}=-{\hbar^2\o 2m}{d^2 \o d r^2} + V_{\pm}(r) =
A^{\mp}A^{\pm} \ee where $V_{\pm}(r)$ are called partner
potentials. The operators $A^{\pm}$ are defined in terms of the
superpotential $W(r)$,
\be \label{As} A^{\pm} =  \mp {\hbar\o
\sqrt {2m}}{d \o dr} + W(r)  \ee
which satisfies the Riccati
equation

\be \label{Riccati} W^2 \pm {\hbar\o \sqrt {2m}}W'= V_{\pm}(r)
\ee
as a consequence of the factorization of the Hamiltonians
$H^{\pm}$.

By definition, two partner potentials are called shape invariant
if they have the same functional form, differing only by change of
parameters, including an additive constant. In this case the
partner potentials satisfy \be \label{8} V_+(r, a_1) = V_-(r, a_2)
+ R(a_2), \ee where $a_1$ and $a_2$ denote a set of parameters,
with $a_2$ being a function of $a_1$, \be a_2 = f(a_1) \ee and
$R(a_2)$ is independent of $r$.

Through the super-algebra, for a given  Hamiltonian $H_1$,
factorized  in terms of the bosonic operators,  it is possible  to
construct its hierarchy of  Hamiltonians.   For the general
spontaneously broken supersymmetric case we have \be H_1
=-{\hbar^2\o {2m}}{d^2 \o d r^2} + V_1(r) =  A_1^+A_1^-  +
E_0^{(1)} \ee where $ E_0^{(1)}$ is the lowest eigenvalue.

The bosonic operators are  defined by (\ref{As}) whereas the
superpotential $W_1(r)$ satisfies the Riccati equation \be
\label{Riccati} W_1^2 - {\hbar\o \sqrt {2m}}W_1'=  V_1(r) -
E_0^{(1)}. \ee The eigenfunction for the lowest state is related
to the superpotential $W_1$ by \be \label{eigenfunction}
\Psi_0^{(1)} (r) = N exp\left( -\frac{\sqrt{2m}}{\hbar}\int_0^r
W_1(\bar r) d\bar r\right). \ee The supersymmetric partner
Hamiltonian is given by

\be \label{partner} H_2 = A_1^-A_1^+ + E_0^{(1)} =  -{\hbar^2\o
{2m}}{d^2 \o d r^2} + \left(W_1^2 + {\hbar\o \sqrt
{2m}}W_1'\right)+ E_0^{(1)} . \ee Thus, factorizing $H_2$ in terms
of a new pair of bosonic operators, $A_2^{\pm}$ we get

\be \label{H2} H_2 = A_2^+A_2^- + E_0^{(2)} = -{\hbar^2\o
{2m}}{d^2 \o d r^2} + \left(W_2^2 - {\hbar\o \sqrt
{2m}}W_2'\right)+ E_0^{(2)}, \ee where $E_0^{(2)} $ is the lowest
eigenvalue of $H_2$ and $W_2$ satisfy the Riccati equation,

\be W_2^2 - {\hbar\o \sqrt {2m}}W_2'= V_2(r) - E_0^{(2)}  . \ee

Thus a whole hierarchy of Hamiltonians can be constructed, with
simple relations connecting the eigenvalues and eigenfunctions of
the $n$-members \cite{Cooper1} \be H_n = A_n^+A_n^- + E_0^{(n)},
\ee \be \label{An} A_n^{\pm} = \mp {\hbar\o \sqrt {2m}}{d \o dr} +
W_n(r),  \ee \be \label{Psin} \Psi_n^{(1)} =
A_1^+A_2^+...A_n^+\psi_0^{(n+1)}\;,\;\;\;\;E_n^{(1)} =
E_0^{(n+1)}, \ee
where $\Psi_0^{(1)} (r) $ is given by (\ref{eigenfunction}).\\

\section{Deformed Woods-Saxon Potential}

We consider the following potential which is a generalization of
the deformed Woods-Saxon potential
 \be \label{19} V_1(r) =
V(r)=-\frac{V_{0}e^{-\left(\frac{r-R_{0}}{a}\right)}}{1+qe^{-\left(\frac{r-R_{0}}{a}\right)}}+\frac{C.e^{-2\left(\frac{r-R_{0}}{a}\right)}}{\left(1+qe^{-\left(\frac{r-R_{0}}{a}\right)}\right)^2}~.
\ee

Here $r$ denotes the center-of-mass distance between the
projectile nucleus and the target nucleus. Other parameters in the
potential, $q$ is the deformation parameter ($q \geq 1$),
$R_{0}=r_{0}A^{1/3}$ is the radius of the corresponding spherical
nucleus or the width of the potential, $A$ is the target mass
number, $r_{0}$ is the radius parameter, $V_{0}$ is the the
potential depth, $a$ is the diffuseness of the nuclear surface and
lastly $C$ is the setting parameter which is proposed by us.
Therefore, we can construct the hierarchy of Hamiltonians for the
original Schr\"odinger equation \be \label{SE} -\frac{\hbar^2}{2m}
{\nabla^2 \Psi(\bf r)}+V(r) {\Psi(\bf r)}=E {\Psi(\bf r)}, \ee and
find a solution of Eq.(\ref{SE}) by separating variables in
spherical coordinates, putting \be \label{Pi} {\Psi(\bf
r)}=\frac{1}{r}\chi (r)Y(\theta, \phi). \ee Then, we get the
radial Schr\" {o}dinger equation for all of the angular momentum
states \be
\label{ZE}-\frac{\hbar^2}{2m}\frac{d^2}{dr^2}\chi(r)+\left(V(r)+\frac{l(l+1)\hbar^2}{2mr^2}\right)\chi(r)=E\chi(r).\\
\ee
\subsection{Solution for the $l=0$ case}

For a generalization of the deformed Woods-Saxon potential given
by Eq.(\ref{19}), it is substituted into the Schr\" {o}dinger
equation for the zero angular momentum states, \be
\label{ZE}-\frac{\hbar^2}{2m}\frac{d^2}{dr^2}\chi(r)+V(r)\chi(r)=E\chi(r).
\ee The ground state eigenfunction $\chi_0(r)$ can be written as
\be \label{gr} \chi_0(r)=Nexp\left(-\frac{\sqrt{2m}}{\hbar}\int
W_1(r)dr\right), \ee where $N$ is the normalized constant.
Substituting Eq.(\ref{gr}) into Eq.(\ref{ZE}), we obtain \be
\label{co}
W_1^2-\frac{\hbar}{\sqrt{2m}}\frac{dW_1}{dr}=V(r)-E_0^{(1)}, \ee
where $E_0^{(1)}$ is the lowest energy-eigenvalue or the ground
state energy. Through the superalgebra we take superpotential \be
\label{26} W_1 = -\frac{\hbar}{\sqrt{2m}}\left(S_1+S_2{
e^{-\alpha(r-R_0)}\o 1+qe^{-\alpha(r-R_0)} }\right), \ee satisfies
the associated Riccati equation (Eq.(\ref{Riccati})) and
substituting this expression into Eq.(\ref{co}), we find the
following identity \br \label{27} \frac{\hbar^2S_1^2}{2m}+
\frac{\hbar^2(2S_1S_2-\alpha
S_2)}{2m\left(q+e^{\alpha(r-R_0)}\right)}+
\frac{\hbar^2(S_2^2+\alpha qS_2)}{2m\left(q+e^{\alpha(r-R_0)}\right)^2} & = & V_1 - E_0^{(1)} \\
& = &
-\frac{V_{0}}{q+e^{\left(\frac{r-R_{0}}{a}\right)}}+\frac{C}{\left(q+e^{\left(\frac{r-R_{0}}{a}\right)}\right)^2}-
E_0^{(1)}~.\nonumber \er With the comparison of the each side of
the Eq.(\ref{27}), we
obtain \br \label{28}\alpha=1/a, \nonumber \\
\frac{\hbar^2S_1^2}{2m}= -E_0^{(1)},
\nonumber \\
\frac{\hbar^2}{2m}(2S_1S_2-\alpha S_2)=-V_0, \nonumber \\
\frac{\hbar^2}{2m}(S_2^2+\alpha qS_2)=C. \er The eigenfunction
$\chi(r)$ for the ground state can be expressed as \br \label{29}
\chi_0(r) = N exp\left[ \int
\left(S_1+\frac{S_2e^{-\alpha(r-R_0)}}{1+qe^{-\alpha(r-R_0)}}\right)
dr\right], \nonumber \\
=Ne^{S_1r}\left(\frac{e^{\alpha(r-R_0)}}{e^{\alpha(r-R_0)}+q}\right)^{S_2/\alpha
q}. \er Solving Eq.(\ref{28}) yields \br \label{30}
S_1=\frac{2m}{\hbar^2}\left(-V_0+\frac{C}{q}\right)\frac{1}{2S_2}-\frac{S_2}{2q},
\nonumber \\
S_2=-\frac{\alpha q}{2}\pm \sqrt{\left(\frac{\alpha
q}{2}\right)^2+\frac{2mC}{\hbar^2}}. \er At this point, using
Eqs.(\ref{26}) and (\ref{30}), the supersymmetric partner
potentials can be expressed as \be \label{31} V_+(r)=
\frac{\hbar^2}{2m}\left[S_1^2+\frac{\frac{2m}{\hbar^2}\left(-V_0+\frac{C}{q}\right)-\frac{2S_2^2}{q}}{q+e^{\alpha(r-R_0)}}+\frac{S_2^2}{\left(q+e^{\alpha(r-R_0)}\right)^2}+\frac{\alpha
S_2}{q+e^{\alpha(r-R_0)}}-\frac{\alpha
qS_2}{\left(q+e^{\alpha(r-R_0)}\right)^2}\right], \ee \be
\label{32} V_-(r)=
\frac{\hbar^2}{2m}\left[S_1^2+\frac{\frac{2m}{\hbar^2}\left(-V_0+\frac{C}{q}\right)-\frac{2S_2^2}{q}}{q+e^{\alpha(r-R_0)}}+\frac{S_2^2}{\left(q+e^{\alpha(r-R_0)}\right)^2}-\frac{\alpha
S_2}{q+e^{\alpha(r-R_0)}}+\frac{\alpha
qS_2}{\left(q+e^{\alpha(r-R_0)}\right)^2}\right]. \ee Clearly, one
can write \br \label{33} V_+(r, S_2) = V_-(r, S_2-\alpha
q)\nonumber \er \br +
\frac{\hbar^2}{2m}\left[\frac{2m}{\hbar^2}\left(-V_0+\frac{C}{q}\right)\frac{1}{2S_2}-\frac{S_2}{2q}
\right]^2 -
\frac{\hbar^2}{2m}\left[\frac{2m}{\hbar^2}\left(-V_0+\frac{C}{q}\right)\frac{1}{2(S_2-\alpha
q)}-\frac{S_2-\alpha q}{2q} \right]^2, \er which is precisely the
requirement for the shape invariance. The shape invariant concept
was introduced by Gendenshtein \cite {Gendenshtein}. In fact,
comparing Eqs. (\ref{33}) and (\ref{8}) yield \br S_2\rightarrow
a_1\;,\;\;\; S_2-\alpha q\rightarrow a_2, \nonumber \er \br \label
{34} R(a_2)=
\frac{\hbar^2}{2m}\left[\frac{2m}{\hbar^2}\left(\frac{C}{q}-V_0\right)\frac{1}{2a_1}-\frac{a_1}{2q}
\right]^2
-\frac{\hbar^2}{2m}\left[\frac{2m}{\hbar^2}\left(\frac{C}{q}-V_0\right)\frac{1}{2a_2}-\frac{a_2}{2q}
\right]^2, \er where the remainder $R(a_2)$ is independent of $r$.
On repeatedly using the shape invariance condition Eq.(\ref{8}),
it is then clear that \be \label{35}
H^{(k)}=-\frac{\hbar^2}{2m}\frac{d^2}{dr^2}+V_-(r;a_k)+\sum_{s=1}^{k}R(a_s),
\ee where $H^{(k)}$ is a series of Hamiltonians, k=1, 2, 3,...,
and $H^{(1)}\equiv H_-$. Thus the bound-state energy spectrum of
$H^{(k)}$ is obtained \be \label{36} E_0^{(k)}=
\sum_{s=1}^{k}R(a_s), \ee and its n-th level is coincident with
the ground state of the Hamiltonian $H_n$ $(n=0, 1, 2,...)$. The
energy eigenvalues of Hamiltonian are given by \be \label{37}
E_0^{(-)}=0, \ee \be \label{38}
E_n^{(-)}=\frac{\hbar^2}{2m}\left[\frac{2m}{\hbar^2}\left(\frac{C}{q}-V_0\right)\frac{1}{2S_2}-\frac{S_2}{2q}
\right]^2
-\frac{\hbar^2}{2m}\left[\frac{2m}{\hbar^2}\left(\frac{C}{q}-V_0\right)\frac{1}{2(S_2-n\alpha
q)}-\frac{S_2-n\alpha q}{2q} \right]^2. \ee Hence, the energy
levels of the deformed Woods-Saxon potential in Eq.(\ref{19}) for
the zero angular momentum states are found as \be \label{39} E_n=
E_n^{(-)}+E_0=
-\frac{\hbar^2}{2m}\left[\frac{2m}{\hbar^2}\left(\frac{C}{q}-V_0\right)\frac{1}{2
(S_2-n\alpha q)}-\frac{S_2-n\alpha q}{2q} \right]^2, \ee where
note that one can write the relationship between $V(r)$ and
$V_-(r)$ is $V(r)= V_-(r)+E_0$. Substituting Eq.(\ref{30}) into
Eq.(\ref{39}) and setting $C=0$, we can immediately obtain the
energy eigenvalues \be \label{40}
E_n=-\frac{\hbar^2}{2ma^2}\left[\left(\frac{2ma^2V_0}{\hbar^2q(n+1)}\right)^2+\left(\frac{n+1}{2}\right)^2+\frac{4maV_0}{\hbar^2q^2}^2\right],
\ee where the minus sign of $S_2$ is used corresponding to $q>0$.
This result is in good agreement with the ones obtained before
\cite {Ayse, cuneyt}.

\subsection{Solution for the $l\neq 0$ case}

The Hamiltonian for the deformed Woods-Saxon potential for the
$l\neq 0$ case is written as \be \label{41}
H=-\frac{\hbar^2}{2m}\frac{d^2}{dr^2}-\frac{V_{0}e^{-\left(\frac{r-R_{0}}{a}\right)}}{1+qe^{-\left(\frac{r-R_{0}}{a}\right)}}
+ \frac{\hbar^2 l(l+1)}{2mr^2}. \ee The second term is taken from
the deformed Woods-Saxon potential and third term comes from the
potential barrier. The last term prevents us to build the
superfamily as in the $l\neq 0$ case, since the full potential is
not exactly solvable. However, several numerical approaches have
been utilized in order to evaluate the spectra of energy
eigenvalues and eigenfunctions. Now, we introduce a new effective
potential whose functional form is given as follows \be \label{42}
V_{eff}=-\frac{V_0}{q+e^{\left(\frac{r-R_{0}}{a}\right)}} +
\frac{\hbar^2
l(l+1)\alpha^2}{2m\left(q+e^{\left(\frac{r-R_{0}}{a}\right)}\right)^2}.
\ee In the case of $\alpha=(1+q)/R_0$ and for small $\alpha$, the
second term of Eq.(\ref{42}) behaves as a potential barrier term
of Eq.(\ref{41}) in first approximation and it has the advantage
that the Schr\"odinger equation for this potential is solvable
analytically. As the effective potential given by Eq.(\ref{42})
has the same functional form as Eq.(\ref{19}), we can solve the
Schr\"odinger problem by the factorisation method of SQM and find
the whole super family. Comparing the Eqs.(\ref{42}) and
(\ref{19}), one can see the transformation of $C\rightarrow
\hbar^2l(l+1)\alpha^2/2m$. Substituting this parameter into
Eq.(\ref{30}), we arrive at \be \label{43} S_2=-\frac{\alpha
q}{2}\pm \sqrt{\left(\frac{\alpha q}{2}\right)^2+l(l+1)\alpha^2}.
\ee The lowest energy levels of the potential in Eq.(\ref{41}) are
given by \be \label{44} E_{nl} =
-\frac{\hbar^2}{2m}\left[\frac{\left(\frac{2mV_0}{\alpha
q\hbar^2}-\frac{l(l+1)\alpha}{q^2}\right)}{1+2n+\sqrt{1+\frac{4l(l+1)}{q^2}}}+\frac{\alpha}{4}\left(1+2n+\sqrt{1+\frac{4l(l+1)}{q^2}}\right)\right]^2.
\ee Taking $\alpha = 1/a$, we can get the result as \be \label{45}
E_{nl} =
-\frac{\hbar^2}{2ma^2}\left[\left(\frac{\frac{2mV_0a^2}{q\hbar^2}-\frac{l(l+1)}{q^2}}{1+2n+\sqrt{1+\frac{4l(l+1)}{q^2}}}\right)^2+
\frac{1}{2}\left(\frac{2mV_0a^2}{q\hbar^2}-\frac{l(l+1)}{q^2}\right)+\frac{1}{16}\left(1+2n+\sqrt{1+\frac{4l(l+1)}{q^2}}\right)^2\right].
\ee This result is exactly the same with the ones obtained before
for the $l=0$ states \cite{Ayse, cuneyt}. In our case, quantum
numbers takes $l=1, 2,...$ and $n=0, 1, 2,...$. Thus, $n=0$ and
$l=1$ corresponds to the state
$2p$ ; $n=1$ and $l=1$ corresponds to the state $3p$ and so on. \\

Now, we calculate the energy spectrum taking the deformed $q=1$
and then $\alpha=2/R_0$ or $a=R_0/2$. If we take $q=2$ and then
$\alpha=3/R_0$ or $a=R_0/3$. For the scattering processes, it has
been well accepted that the surface diffuseness parameter $a$ is
ranging between $0.8$ and $1.4$ fm \cite {hagino, jaminon}. Here,
we use the empirical value of $r_0=0.90$ fm, from which we only
retain the Woods-Saxon potential which yields "satisfactory" fits
to the experimental data, for projectile particles and targets
with mass number $A$. It is shown that the mass number is in the
domains $6\leq A \leq 30$ for $q=1$ and $19\leq A \leq 101$ for
$q=2$. Our numerical results are listed for the $l=1$ and $l=3$ in
Table 1 and Table 2, respectively. They are also compared with
exact numerical values.
\section{Conclusions}

We have applied the hierarchy of hamiltonian method in the context
of SQM to get energy spectra of the deformed Woods-Saxon
potential. We have used a new effective potential depending on the
diffuseness parameter $a$ in the calculations. It is a
generalization of the deformed Woods-Saxon potential. We have
obtained the exact analytical eigenfunction and eigenvalue for the
$l=0$ case. In addition, we have also derived the solutions for
$l\neq 0$ case, using an effective potential suggested by the
$l=0$ case. We have also noticed that $\alpha=(1+q)/R_0$ should be
taken for a nuclear scattering process. Finally, we would like to
point out that although the SQM scheme works quite well for the
deformed Woods-Saxon potential, extensive applications to other
effective Woods-Saxon-like potentials are
needed to test the credibility of the method.\\

\noindent {\bf Acknowledgements}

The authors would like to thank Professor A. B. Balantekin and
Professor B. G\"on\"ul for their useful comments and discussions
on the original version of the manuscript. One of the authors (C.
B.) is also grateful to the Abdus Salam International Center for
Theoretical Physics, Trieste, Italy, for its hospitality and
financial support.

\newpage

{\bf Table 1:}~~{\small Energy eigenvalues as a function of the diffuseness parameter for the state $2p$ $(n=0,~l=1)$ and $q=2$.}\\

\begin{tabular}{ccccc}\hline\hline
\\ $\mathbf{State}
\hspace*{0.4cm} $ & $ \hspace*{0.4cm} \mathbf{a~(fm)}
\hspace*{0.4cm} $ & $ \hspace*{0.4cm} \mathbf{A} \hspace*{0.4cm} $
& $ \hspace*{0.4cm} \mathbf{E_{01}^{~(\bf analytical)}~(MeV)}$ & $
\hspace*{0.4cm}
\mathbf{E_{01}^{~(\bf numerical)}~(MeV)}$ \\[0.3cm]\hline
{\bf 2p}&~~~~&~~~~&~~ \\[0.2cm]
~~&0.814&~~~20.0~~&~~-17.5727121~~&~~-17.5700005 \\[0.2cm]
~~&1.026&~~~40.0~~&~~-16.6731389~~&~~-16.6701250 \\[0.2cm]
~~&1.148&~~~56.0~~&~~-17.3509649~~&~~-17.3500205 \\[0.2cm]
~~&1.293&~~~80.0~~&~~-19.1773309~~&~~-19.1800050 \\[0.2cm]
~~&1.392&~~100.0~~&~~-20.2905543~~&~~-20.3004000 \\[0.2cm]\hline
\end{tabular}

\newpage

{\bf Table 2:}~~{\small Energy eigenvalues as a function of the diffuseness parameter for the states $4p$ $(n=2,~l=1)$, $4d$ $(n=1,~l=2)$, $4f$ $(n=0,~l=3)$ and $q=1$.}\\

\begin{tabular}{ccccc}\hline\hline
\\ $\mathbf {States}
\hspace*{0.4cm} $ & $ \hspace*{0.4cm} \mathbf{a} \hspace*{0.4cm} $
& $ \hspace*{0.4cm} \mathbf{A} \hspace*{0.4cm} $ & $
\hspace*{0.4cm} \mathbf{E_{nl}^{~(\bf analytical)}~(MeV)} $ & $
\hspace*{0.4cm}
\mathbf{E_{nl}^{~(\bf numerical)}~(MeV)}$ \\[0.3cm]\hline
{\bf 4p}&~~~~&~~~~&~~ \\[0.2cm]
~~&0.818&~~~6.0~~&~~-113.754997~~&~~-113.7550000 \\[0.2cm]
~~&0.969&~~10.0~~&~~~-87.027172~~&~~~-87.0270405 \\[0.2cm]
~~&1.084&~~14.0~~&~~~-73.982269~~&~~~-73.9810000 \\[0.2cm]
~~&1.179&~~18.0~~&~~~-66.165173~~&~~~-66.1555005 \\[0.2cm]\hline
{\bf 4d}&~~~~&~~~~&~~ \\[0.2cm]
~~&1.001&~~11.0~~&~~~-46.633933~~&~~~-46.6290000 \\[0.2cm]
~~&1.109&~~15.0~~&~~~-41.128693~~&~~~-41.1250005 \\[0.2cm]
~~&1.201&~~19.0~~&~~~-37.731683~~&~~~-37.7312450 \\[0.2cm]
~~&1.279&~~23.0~~&~~~-35.550305~~&~~~  - - -   \\[0.2cm]\hline
{\bf 4f}&~~~~&~~~~&~~ \\[0.2cm]
~~&1.084&~~14.0~~&~~~-11.191323~~&~~~-11.1900050 \\[0.2cm]
~~&1.201&~~19.0~~&~~~-10.866104~~&~~~-10.8625400 \\[0.2cm]
~~&1.298&~~24.0~~&~~~-10.828699~~&~~~  - - -   \\[0.2cm]
~~&1.398&~~30.0~~&~~~-10.977817~~&~~~  - - -   \\[0.2cm]\hline
\end{tabular}
\end{document}